\documentclass[conference]{IEEEtran}
\IEEEoverridecommandlockouts
\usepackage{cite}
\usepackage{amsmath,amssymb,amsfonts}
\usepackage{algorithmic}
\usepackage{graphicx}
\usepackage{textcomp}
\usepackage{xcolor}
\usepackage{url}
\usepackage{booktabs}
\def\BibTeX{{\rm B\kern-.05em{\sc i\kern-.025em b}\kern-.08em
    T\kern-.1667em\lower.7ex\hbox{E}\kern-.125emX}}
\begin{document}

\title{A Rule-Aware Prompt Framework for Structured Numeric Reasoning in Smart Grids\\
}

\author{\IEEEauthorblockN{Yichen Liu, Hongyu Wu}
\IEEEauthorblockA{\textit{Dept. of Electrical and Computer Engineering} \\
\textit{Kansas State University}\\
Manhattan, KS, USA \\
\{ccwudi, hongyuwu\}@ksu.edu}
\and
\IEEEauthorblockN{Bo Liu}
\IEEEauthorblockA{\textit{School of Engineering and Applied Sciences} \\
\textit{Washington State University}\\
Richland, WA, USA \\
bo.liu1@wsu.edu}
}

\maketitle

\begin{abstract}
Smart grids rely on high-dimensional numeric telemetry and explicit operating rules to maintain reliable and secure operation. Recent large language models (LLMs) are increasingly considered as candidate decision-support components for power system operations, yet most deployments focus on textual logs, alerts, or operator messages and do not directly address rule-grounded reasoning over numeric grid measurements. This paper proposes a rule-aware prompt framework that systematically encodes power system domain context, numeric normalization, and decision rules into a modular prompt architecture for LLMs. The framework decomposes prompts into reusable modules, including role, domain context, numeric normalization, rule-aware reasoning, value block, and output schema, and exposes an interface for plugging in diverse grid operating rules. A key design element separates rule specification from the representation of normalized numeric deviations, enabling concise prompts aligned with power system criteria. To illustrate its behavior, we instantiate the framework on numeric anomaly detection in the IEEE 118-bus transmission network and evaluate several prompting and adaptation regimes. The results show that rule-aware, z-score-based value blocks and a hybrid LLM+DL architecture substantially improve both consistency with grid operating rules and anomaly detection performance while reducing token usage, providing a reusable bridge between grid telemetry and general-purpose LLMs.
\end{abstract}

\begin{IEEEkeywords}
Prompt engineering, large language models, anomaly detection, power system monitoring, rule-based reasoning
\end{IEEEkeywords}

\section{Introduction}
Modern smart grids are engineered around structured numeric telemetry and explicit operating rules that together sustain reliable and secure electricity delivery. Measurements such as bus voltage magnitudes, nodal active and reactive power injections, and line power flows are collected at high frequency from supervisory control and data acquisition (SCADA) systems and phasor measurement units (PMUs), and are interpreted against statistical control limits, protection thresholds, and physical invariants derived from AC power flow. Extracting actionable decisions from these data typically relies on application-specific estimators, classifiers, and carefully tuned detection pipelines, which can be difficult to adapt as grid topology, operating conditions, and sensing infrastructure evolve.

In parallel, large language models (LLMs) have demonstrated strong capabilities in natural language understanding, code generation, and tool-augmented decision support~\cite{MinaeeLLM2025,HanPretrained2021,HuynhCodeGen2025}. Built primarily on the Transformer architecture~\cite{Vaswani2017}, these models can support diverse tasks through prompting, in-context learning, and lightweight fine-tuning. Recent surveys and position papers highlight the emerging role of LLMs as agentic decision-support tools across domains~\cite{PlaatAgentic2025}. In the power sector, LLMs are being explored for system analysis, market operations, planning, and operator support, suggesting a promising pathway toward agentic intelligence in grid operations~\cite{ZhangPowerAgent2024,MajumderEnergyLLM2024}.

However, most existing grid-oriented deployments remain predominantly text-centric, relying on operator logs, alarm messages, and documentation rather than directly operating on high-dimensional numeric telemetry governed by explicit rules. Anomaly detection (AD), which is critical for grid resilience against sensor faults, measurement errors, and cyber intrusions, is a representative downstream task in which this gap is particularly visible~\cite{ZaboliADGrid2024}. When numeric grid data are naively serialized as tokens, LLM behavior can become sensitive to representation choices and may drift away from the underlying decision rules, even if those rules are available to the model in textual form. This is especially problematic in safety-critical power system operations, where decisions must remain traceable to well-understood statistical or physical criteria~\cite{XuLRMSurvey2025}.

Prior work has extensively studied prompt engineering as a means of steering LLM behavior without modifying model parameters~\cite{AgarwalPromptWizard2024}, alongside parameter-efficient adaptation~\cite{HuLoRA2021} and zero-/few-shot and in-context learning paradigms~\cite{XianZeroShot2019,ParnamiFewShot2022,DongICL2024}. Related studies on prompt quality have emphasized process-level prompt design and user-facing factors~\cite{WilbersOPE2024,ChoiPrompt2025}, and ontology-chatbot pipelines use LLMs as interfaces to formal engineering knowledge~\cite{ReifOntology2024}. Diagnostic LLMs have also shown promise for structured reasoning tasks such as conflict and diagnosis generation~\cite{SztyberDX2025}. However, these efforts operate primarily on textual interfaces; relatively few studies directly address high-dimensional numeric grid telemetry governed by multiple operating rules simultaneously, which is the gap this paper targets.

A central challenge is therefore: How should prompts be structured so that LLMs can reason over numeric grid telemetry in a way that is faithful to power system operating rules, interpretable to grid operators, and efficient in token usage? Addressing this challenge requires more than selecting a particular backbone model or tuning strategy. It calls for a systematic way to encode sensor semantics, normalization, and rule logic into prompts that can be reused across grid monitoring tasks such as anomaly detection, bad-data identification, and event classification.

Motivated by these observations, we propose a \emph{rule-aware prompt framework} for structured numeric reasoning in smart grids. Instead of treating grid telemetry as generic tokens, the framework encodes power system domain context, measurement semantics, and decision rules into a modular prompt architecture that accommodates zero-shot prompting, in-context learning, parameter-efficient fine-tuning, and hybrid LLM--deep-learning detectors within a single structure. A key design choice is the explicit separation of (i) a module that specifies how raw grid measurements are normalized into descriptors suitable for rule-based reasoning and (ii) a module that encodes the operating rules themselves. This separation enables concise value blocks that are tightly aligned with grid decision criteria while remaining agnostic to the choice of LLM backbone. Rather than replacing a conventional threshold-based detector, the LLM here integrates explicit rules, normalized multivariate evidence, and structured output constraints within a unified reasoning interface that also yields natural-language rationales, with anomaly detection serving as one illustrative grid monitoring task rather than the primary contribution.

Concretely, each prompt is composed of six modules: role instruction ($R$), domain context ($C$), numeric normalization ($N$), rule-aware reasoning ($S$), value block ($V$), and output schema ($O$). While we instantiate $S$ with a standard three-sigma statistical rule in our case study, the design is intended to accommodate richer power system rule sets, including physical invariants such as AC power-flow consistency and protection-related operating limits, without changing the overall architecture.

To demonstrate how this framework behaves in practice, we apply it to numeric anomaly detection in a transmission network, using GPT-OSS-20B as a representative open-weight LLM. Within this case study, we examine how different normalization strategies and supervision levels (e.g., zero-/few-shot prompts, in-context learning, lightweight adaptation, and hybrid integration with traditional detectors) interact with the framework to affect rule adherence, interpretability, and token efficiency. The main contributions are threefold:

\begin{itemize}
    \item A modular, rule-aware prompt framework for numeric grid telemetry, in which role, domain context, numeric normalization, rule specification, value block, and output schema are explicitly separated.
    \item A principled formulation of rule-grounded prompt design that combines normalized descriptors with explicit statistical rules so that LLM decisions are based on deviations aligned with power system operating criteria rather than raw measurements.
    \item An IEEE 118-bus case study that quantitatively analyzes how input normalization and supervision level affect detection performance, rule adherence, interpretability, and token efficiency, and examines compatibility with hybrid LLM+DL pipelines for grid monitoring.
\end{itemize}

\section{Rule-Aware Prompt Framework}
\label{sec:framework}
This section presents the proposed prompt framework for structured numeric reasoning in smart grids. The design aims to transform multivariate grid telemetry into a textual format that (i) exposes power system measurement semantics, (ii) encodes explicit operating rules, (iii) maintains a concise and regular structure amenable to LLM inference, and (iv) is composed of reusable modules that can be recombined across grid monitoring applications with minimal changes.

\subsection{Modular Architecture}
The proposed framework decomposes a prompt into six modules:
\begin{itemize}
    \item \textbf{Role Instruction Module (R):} Defines the model's role and high-level task, e.g., ``You are a power system analyst. Your task is to assess whether the following grid telemetry is consistent with normal operation according to a given statistical rule.''
    \item \textbf{Domain Context Module (C):} Provides a concise description of the power system under study, measurement types, and feature semantics. This helps the LLM associate numeric quantities with physical meaning, such as bus voltage magnitudes, nodal active and reactive power injections, and line power flows.
    \item \textbf{Numeric Normalization Module (N):} Specifies how raw grid measurements are transformed into normalized quantities, such as z-scores relative to nominal operating statistics, to reduce scale sensitivity across heterogeneous electrical quantities and limit token redundancy.
    \item \textbf{Rule-Aware Reasoning Module (S):} Encodes the decision rule (e.g., the three-sigma criterion commonly used for bad-data detection in grid monitoring) and the required reasoning steps. The LLM is instructed to apply the rule explicitly when forming its judgment.
    \item \textbf{Value Block Module (V):} Presents the grid telemetry in a compact textual form (e.g., a table or list of normalized descriptors aligned with the rule), as detailed in Section~\ref{sec:valueblock}.
    \item \textbf{Output Schema Module (O):} Constrains the response to a small number of fields, such as a categorical label and a short explanation, for downstream use by operators or automated pipelines.
\end{itemize}

A generic prompt instance can be represented as
\begin{equation}
    P = R \oplus C \oplus N \oplus S \oplus V \oplus O,
\end{equation}
where $V$ denotes the value block containing the numeric telemetry representation, and $\oplus$ denotes concatenation. In the case study, $R$, $C$, $S$, and $O$ are held fixed across experiments, while $N$ and the design of $V$ are varied to assess the impact of different numeric representations.

\subsection{Three-Sigma Statistical Rule}
To provide a concrete and interpretable rule for numeric reasoning in grid monitoring, we use the three-sigma criterion, a widely adopted statistical baseline for detecting outliers in multivariate measurement streams. Let $x_i$ denote the value of the $i$-th grid measurement, and let $\mu_i$ and $\sigma_i$ denote its nominal mean and standard deviation estimated from historical operating data. The normalized z-score is
\begin{equation}
    z_i = \frac{x_i - \mu_i}{\max(\sigma_i, \epsilon)},
\end{equation}
where $\epsilon$ is a small positive constant to avoid division by zero. A simple rule for flagging deviations is
\begin{equation}
    \text{flag}_i =
    \begin{cases}
        1, & |z_i| \geq \tau,\\
        0, & \text{otherwise},
    \end{cases}
\end{equation}
where $\tau$ is a threshold, commonly set to $3.0$ in three-sigma control charts.

In the prompt, we encode the rule in natural language and provide normalized values (e.g., absolute z-scores) in the value block. The LLM is instructed to determine whether the overall grid snapshot is nominal or abnormal and to briefly justify its decision. Alternative rule modules (e.g., robust statistics, physical residual checks, or protection-limit constraints) can be substituted to handle non-Gaussian operating regimes without changing the overall framework.

\subsection{Value Block Representation}
\label{sec:valueblock}
The value block $V$ presents the grid telemetry in a tabular or list format, with each row corresponding to a single sensor channel (e.g., a bus voltage magnitude or a line flow). To manage token length across the hundreds of measurements contained in a modern transmission snapshot, we focus on concise numeric descriptors and identifiers. Four concrete variants are evaluated in the case study (see Section~\ref{sec:experiments}); the \emph{z-score-only} design corresponds directly to the normalized deviation representation emphasized by the framework and is the default configuration used in all non-ablation experiments. By omitting raw values and statistics and relying instead on absolute z-scores aligned with the decision rule, this representation substantially reduces token count while preserving the information required for rule-based reasoning on grid data.

\subsection{Output Schema}
To facilitate consistent and machine-actionable responses, the output schema module $O$ prescribes a simple format:
\begin{enumerate}
    \item A single-word label indicating the overall assessment, such as ``normal'' or ``anomaly''.
    \item A short explanation summarizing which measurements, if any, exceeded the threshold and why the rule implies the chosen label.
\end{enumerate}
By constraining the output, we reduce variability and simplify downstream integration with grid monitoring tools, while still allowing the model to provide interpretable rationales that operators can inspect.

\section{Experimental Setup}
\label{sec:experiments}
This section instantiates the proposed rule-aware prompt framework on a numeric anomaly detection task in a representative transmission network. The goal is to assess how the modular prompt design, and in particular the three-sigma rule and normalized value block, affect detection performance, rule adherence, and token efficiency.

\subsection{Test System and Dataset}
We consider the IEEE 118-bus AC power system as a representative transmission network. The system is instrumented with sensors measuring nodal active and reactive power injections, line power flows, and bus voltage magnitudes. Each telemetry snapshot is represented as a 255-dimensional vector of real-valued measurements, corresponding to
\begin{itemize}
    \item active/reactive power injections $(P,Q)$ at buses,
    \item active/reactive line flows $(P_f,Q_f)$ on transmission lines, and
    \item voltage magnitudes $V$ at buses.
\end{itemize}

Historical operating data are generated by mapping hourly load profiles from the Electric Reliability Council of Texas (ERCOT) to the system load buses~\cite{ERCOT,ACPFData} and solving AC power flow for each operating point. A subset of snapshots is treated as nominal and used to estimate the mean $\mu_i$ and standard deviation $\sigma_i$ for each measurement. Anomalous states are created by injecting a $15\%$ deviation into three randomly selected sensor values per snapshot, following the compromised-state model $x_a = x_i + a$ commonly used in grid AD studies. For the direct three-sigma baseline, a sample is labeled anomalous if $|z_i| \geq 3.0$ for at least one channel; because measurement channels are strongly correlated through the AC power-flow equations, the observed false-alarm rate settles at $22\%$ rather than the much larger value implied by treating the channels as independent. The LLM-based methods are instead evaluated against the dataset ground-truth labels. For supervised training (LoRA), we use $600$ nominal and $600$ anomalous samples; an additional $400$ labeled samples are evenly split into validation and test sets, and all reported metrics are averaged over the held-out test set.

\subsection{LLM Backbone and Prompting Modes}
The framework is instantiated on GPT-OSS-20B, an open-weight mixture-of-experts (MoE) Transformer language model with approximately $21$~billion total parameters and around $3.6$~billion active parameters per token~\cite{GPTOSS20B}. GPT-OSS-20B is pretrained using an autoregressive language modeling objective on a large-scale multilingual and domain-diverse corpus and is publicly available via the Hugging Face platform. This choice allows us to evaluate LLM performance on structured numeric grid AD under realistic resource constraints that are compatible with utility-side computing environments. All experiments are conducted on a workstation equipped with two NVIDIA RTX~6000 Ada GPUs with 48~GB each.

We evaluate the prompt framework under five configurations that represent increasing levels of supervision and adaptation, mirroring a practical deployment path from zero-touch monitoring to utility-tuned detectors:
\begin{itemize}
    \item \textbf{Zero-shot:} The model receives a single grid telemetry snapshot and the rule-aware prompt, but no labeled examples, following the classical notion of zero-shot learning~\cite{XianZeroShot2019}.
    \item \textbf{Few-shot:} The prompt includes two labeled exemplars (one nominal, one anomalous) before the test snapshot, providing minimal supervision in the prompt~\cite{ParnamiFewShot2022}.
    \item \textbf{In-context learning (ICL):} The prompt includes five diverse labeled examples that cover different anomaly patterns observed across grid operating conditions, exploiting the model's ability to infer tasks from in-context examples~\cite{DongICL2024}.
    \item \textbf{LoRA fine-tuning:} Parameter-efficient low-rank adaptation is applied to the attention projections, supervising the model to generate both the anomaly label and a brief rationale~\cite{HuLoRA2021}.
    \item \textbf{Hybrid LLM+DL detector:} The LLM is used as a rule-aware feature selector and explanation engine; its filtered measurement channels are then passed to a traditional deep learning (DL) anomaly detector for the final decision, keeping real-time inference on the classical detector while leveraging the LLM offline.
\end{itemize}

In all configurations, the same modular prompt architecture $P = R \oplus C \oplus N \oplus S \oplus V \oplus O$ is used, as described in Section~\ref{sec:framework}. The role instruction ($R$), domain context ($C$), rule-aware reasoning module ($S$), and output schema ($O$) are held fixed. The primary experimental degrees of freedom are (i) the numeric normalization ($N$) together with the value block ($V$) content and (ii) the level of supervision (zero-/few-shot, ICL, LoRA, hybrid).

\subsection{Value Block Variants and Evaluation Metrics}
To study the effect of the numeric normalization module and the amount of statistical context exposed to the model, we consider four value-block designs:
\begin{enumerate}
    \item \textbf{Setup 1 (Value only):} Each sensor is represented only by its raw measurement value.
    \item \textbf{Setup 2 (Mean + std + value):} Each row contains the raw value together with its nominal mean and standard deviation from historical grid data.
    \item \textbf{Setup 3 (Mean + std + value + z-score):} Each row additionally includes the computed z-score.
    \item \textbf{Setup 4 (Z-score only, proposed):} Each sensor is represented only by its absolute z-score $\lvert z_i \rvert$, as in the framework description in Section~\ref{sec:valueblock}.
\end{enumerate}
All four variants share the same role, context, and rule description; only the numeric representation in $V$ changes.

We report standard classification metrics: accuracy (overall correctness), recall (fraction of true grid anomalies correctly detected), precision (fraction of predicted anomalies that are correct), and F1-score (harmonic mean of precision and recall). Since each method is asked to apply the encoded three-sigma rule to the same normalized inputs, these metrics jointly reflect detection performance and adherence to the grid operating rule. In addition, we report token counts to compare prompt compactness across value-block designs, which directly impacts inference cost and scalability to larger grids.

\section{Results and Discussion}
This section reports the empirical performance of the rule-aware prompt framework on the IEEE 118-bus case study. We first present an ablation study on value-block design in the zero-shot setting, then compare different prompting and adaptation regimes, and finally assess a hybrid LLM+DL detector. Throughout, the \emph{z-score-only} value block represents the proposed normalized design, and the analysis focuses explicitly on how the prompt architecture influences rule-grounded behavior on grid telemetry.

\begin{figure}[t]
    \centering
    \includegraphics[width=\linewidth]{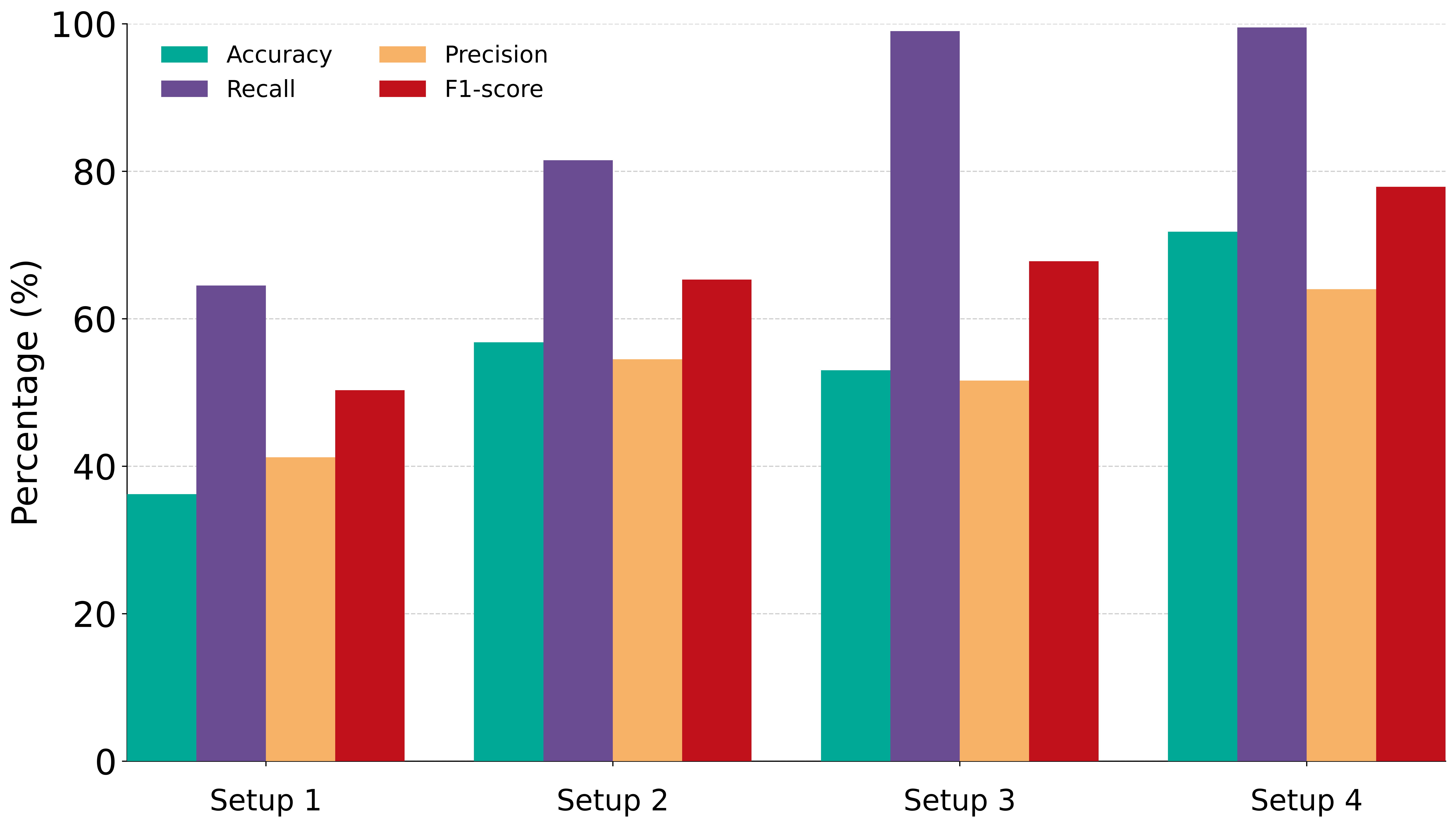}
    \caption{Zero-shot anomaly detection performance under different value-block designs. All configurations share the same role, context, and three-sigma decision rule; only the numeric representation in the value block is changed.}
    \label{fig:ablation_valueblock}
\end{figure}

\subsection{Zero-Shot Performance Under Different Value-Block Designs}

Figure~\ref{fig:ablation_valueblock} summarizes the zero-shot results for four
value-block configurations. Using raw values only (Setup~1) yields limited
performance: accuracy is $36.2\%$ and the F1-score reaches $50.3\%$, indicating that
the model struggles to infer stable thresholds from raw telemetry. Introducing the mean and standard deviation (Setup~2) noticeably improves performance,
achieving $56.8\%$ accuracy and an F1-score of $65.3\%$. Adding z-scores alongside
raw values (Setup~3) further increases recall to $99.0\%$ and raises the F1-score to
$67.8\%$, suggesting that rule-aligned normalization aids the model's zero-shot
reasoning. The best results occur when the value block contains only z-scores (Setup~4), achieving $71.8\%$ accuracy and a $77.9\%$ F1-score. These results indicate that the structure and precision of the value block matter more than its size: presenting essential numeric cues in a concise, rule-aligned form outperforms simply exposing more measurements.

\subsection{Effect of Supervision and Adaptation}
We next fix the value block to the proposed z-score-only representation and compare
five prompting and adaptation regimes: zero-shot, few-shot, in-context learning
(ICL), LoRA fine-tuning, and a hybrid LLM+DL detector. We also include a direct three-sigma thresholding baseline on the same normalized inputs, since the underlying decision rule can be executed without an LLM and therefore provides a useful reference point for interpreting the role of LLM-based reasoning. The results on the held-out test set are reported in Table~\ref{tab:prompting}.

\begin{table}[t]
\centering
\caption{Performance of LLM-Based Methods for AD Compared with a Three-Sigma Baseline}
\label{tab:prompting}
\begin{tabular}{lcccc}
\toprule
\textbf{Prompt paradigm} & \textbf{Acc. (\%)} & \textbf{Rec. (\%)} & \textbf{Prec. (\%)} & \textbf{F1 (\%)} \\
\midrule
Zero-shot  & 71.8 & 99.5 & 64.0 & 77.9 \\
Few-shot   & 79.0 & 84.0 & 76.4 & 80.0 \\
ICL        & 80.5 & 86.0 & 77.4 & 81.5 \\
LoRA fine-tuning & 84.0 & 100 & 75.8 & 86.2 \\
Hybrid LLM+DL & 94.0 & 88.0 & 100 & 93.6 \\
\midrule
3$\sigma$ baseline   & 88.8 & 100 & 81.7 & 89.9 \\
\bottomrule
\end{tabular}
\end{table}

Performance improves steadily across the five LLM-based setups. The zero-shot prompt already
provides a strong starting point, reaching an F1-score of nearly $78\%$ with
high recall. Adding a small number of labeled examples in the few-shot and ICL
regimes improves the balance between true positives and false alarms, raising the F1-score
above $80\%$, especially when examples are diverse~\cite{DongICL2024}.
LoRA fine-tuning further increases recall to $100\%$ but lowers precision,
resulting in an F1-score of $86.2\%$. The hybrid approach achieves the best
overall performance, reaching over $93\%$ F1 with perfect precision.

Among the LLM-based configurations, only the hybrid method surpasses the
$89.9\%$ F1-score of the direct thresholding baseline. This comparison is included
not to argue that either approach is universally superior, but to isolate how
different prompting and adaptation regimes affect rule adherence and interpretability
under a shared statistical criterion.

\subsection{Hybrid LLM+DL Detector}
Table~\ref{tab:hybrid} compares a standalone deep learning detector with the proposed
hybrid model. In the hybrid pipeline, the LLM first analyzes the z-score-based
telemetry under the three-sigma rule and identifies the measurement channels most
relevant to anomaly discrimination; these channels are then passed to the downstream
detector for final classification.

The standalone DL detector achieves high recall ($98.0\%$) but low precision
($80.3\%$), yielding an F1-score of $88.3\%$ due to frequent false alarms, an
unacceptable trade-off in grid operations where each false alarm adds to operator
workload. The hybrid model raises precision to $100\%$ while maintaining $88.0\%$
recall, yielding an F1-score of $93.6\%$. This suggests that the main value of the
LLM lies not in replacing the detector, but in providing rule-aware feature selection
together with interpretable rationales, adding transparency that supports operator
trust in grid monitoring pipelines.

\begin{table}[t]
\centering
\caption{Comparison Between a Traditional DL Detector and the Proposed LLM-Enhanced Hybrid Detector}
\label{tab:hybrid}
\begin{tabular}{lcccc}
\toprule
\textbf{Model} & \textbf{Acc. (\%)} & \textbf{Rec. (\%)} & \textbf{Prec. (\%)} & \textbf{F1 (\%)} \\
\midrule
Traditional DL  & 87.0 & 98.0 & 80.3 & 88.3 \\
Hybrid LLM+DL & 94.0 & 88.0 & 100 & 93.6 \\
\bottomrule
\end{tabular}
\end{table}

\subsection{Token Efficiency and Latency}
Table~\ref{tab:token_reply_time} compares the token usage and response latency of different prompting and adaptation methods. Among the zero-shot setups, token count and reply time increase with prompt complexity, with Setup~3 being the most verbose and slowest to execute. In contrast, Setup~4 achieves higher F1 with a token count comparable to Setup~1 and far smaller than Setups~2 and~3, suggesting that a concise, rule-aligned value representation improves efficiency. Few-shot and ICL methods require significantly more tokens due to labeled examples, with ICL exceeding 11{,}000 tokens. LoRA fine-tuning keeps the token count as low as Setup~4 but incurs the highest latency due to model loading and inference overhead. The hybrid LLM method is the most efficient, with negligible reply time since the LLM is used only for pre-filtering and not during runtime inference.

\begin{table}[t]
\centering
\caption{Comparison of Token Count and Reply Time for Different Methods}
\label{tab:token_reply_time}
\begin{tabular}{lcc}
\toprule
\textbf{Method} & \textbf{Token Count} & \textbf{Reply Time (s)} \\
\midrule
Zero-shot Setup 1      & 2189   & 46.43 \\
Zero-shot Setup 2      & 4287  & 67.88 \\
Zero-shot Setup 3      & 5312  & 68.99 \\
Zero-shot Setup 4      & 2203   & 47.21 \\
Few-shot     & 6320  & 51.73 \\
ICL          & 11461  & 55.99 \\
LoRA fine-tuning  & 2203   & 79.01 \\
Hybrid LLM+DL & -  & $<1$ \\
\bottomrule
\end{tabular}
\end{table}

Across all experiments, performance improves as the value block becomes more structured and concise. Replacing raw values with z-scores (Setup~4) yields a substantially better F1 at a token count comparable to Setup~1 and far smaller than Setups~2 and~3, showing that well-aligned numeric representation is more effective than simply exposing more statistical context. Beyond the value block, prompt paradigms also impact performance. Transitioning from zero-shot to few-shot and in-context learning improves generalization, but increases prompt length and inference cost. LoRA fine-tuning offers higher recall but adds runtime latency, while the hybrid model achieves the best trade-off by using the LLM only for upstream measurement-channel filtering. As shown in Table~\ref{tab:token_reply_time}, prompt complexity and token count directly affect latency, and concise prompts like Setup~4 or the hybrid design offer both accuracy and efficiency.

\section{Conclusion and Future Work}
This paper presented a rule-aware prompt framework for structured numeric reasoning in smart grids, decomposing prompts into modular components (role, domain context, numeric normalization, rule-aware logic, value block, and output schema) so that LLMs can process high-dimensional grid telemetry under explicit power system operating rules. Instantiated on anomaly detection in the IEEE 118-bus transmission system with GPT-OSS-20B, the experiments yield three findings: (i) the proposed z-score-only value block raises zero-shot F1 from roughly $50\%$ to $78\%$ while keeping the prompt compact; (ii) few-shot, in-context learning, and LoRA fine-tuning progressively improve performance, with LoRA achieving very high recall at the cost of added complexity; and (iii) a hybrid LLM+DL architecture attains the best trade-off, exceeding $90\%$ F1 with perfect precision. Clarity and rule alignment in prompt design thus matter more than input volume for grid monitoring tasks. Future work will incorporate richer power system rules such as AC power-flow consistency, protection limits, and topology-aware logical constraints, extend the framework to larger networks and to distribution systems and microgrids with high penetration of distributed energy resources, and evaluate robustness under dynamic operating conditions and coordinated false-data injection attacks.

\section*{Acknowledgment}
The authors would like to thank the U.S. National Science Foundation for its support under Grants Nos. 2146156 and 2316355. 

\section*{AI Use Disclosure}
The authors used AI-assisted language tools solely to improve grammar, clarity, and presentation. All technical content, methodological design, analysis, and conclusions were developed and verified by the authors.

\end{document}